\documentclass[final,3p,times,sort&compress]{elsarticle}

\usepackage{amssymb}

\journal{Nuclear Physics A}

\usepackage{caption}
\usepackage{subcaption}
\usepackage{amsmath}
\usepackage{graphicx}
\usepackage[colorlinks]{hyperref}
\usepackage{float}
\usepackage{tablefootnote}
\usepackage{cleveref}
\usepackage{isotope}
\usepackage[version=4]{mhchem}
\usepackage{multirow}

\begin{document}
\renewcommand{\arraystretch}{1.3}

\begin{frontmatter}

\title{Probing deformed nuclei: Experimental insights into excited states of \isotope[152,153,154]{Gd} isotopes through fusion-evaporation reactions}

\author[first]{S. Pelonis\fnref{label1}}
\author[first]{T.J. Mertzimekis\corref{cor1}}
\ead{tmertzi@phys.uoa.gr}
\author[first]{A. Chalil}
\author[first]{P. Vasileiou}
\author[first]{A. Zyriliou}
\author[first]{G. Zagoraios}
\author[first]{D. Papaioannou}

\author[second]{F.C.L.~Crespi}
\author[second]{A. Bracco}
\author[third]{N. Florea}
\author[third]{N. Marginean}
\author[third]{L. Stan}
\author[third]{A. Turturica}
\affiliation[first]{organization={National and Kapodistrian University of Athens},
            addressline={Zografou Campus},
            city={Athens},
            postcode={GR-15784},
            country={Greece}}
\affiliation[second]{organization={Universita degli Studi di Milano and INFN sez. Milano},
            city={Milano},
            postcode={I-20122},
            country={Italy}}
\affiliation[third]{organization={National Institute for Physics and Nuclear Engineering},
            city={Magurele},
            postcode={R-77125},
            country={Romania}}

\cortext[cor1]{Corresponding author.}
\fntext[label1]{Present address: IKS, KU Leuven, Belgium}

\begin{abstract}
The rare-earth region has been the focus of various studies aiming at the
understanding of nuclear structure and providing information on the
details of the reaction mechanism. The Gd isotopes belong to this group
of nuclei and despite the available spectroscopic information, several
open questions about their structure still exist, such as the inter-band
transitions related to shape evolution or branching-ratios in deformed
states. In addition, production  cross sections of different reactions
to Gd isotopes are largely unknown.

In this work, we report on an experimental attempt to populate the excited
states in the isotopes \isotope[152,153,154]{Gd} by employing the heavy-ion
fusion reaction
\isotope[18]{O}+\isotope[138]{Ba} $\rightarrow$ \isotope[156-x]{Gd} + $x n$
in the 58-64~MeV energy range (center-of-mass). The experiment was conducted
at the 9~MV FV Pelletron Tandem at the Horia Hulubei National Institute for Physics and Nuclear
Engineering, employing the ROSPHERE array. Several branching-ratios for
energy levels in \isotope[152,153]{Gd} have been measured, offering new
and updated values. Furthermore, relative cross sections regarding the
fusion-evaporation reactions
\isotope[138]{Ba}(\isotope[18]{O}, 4n)\isotope[152]{Gd},
\isotope[138]{Ba}(\isotope[18]{O}, 3n)\isotope[153]{Gd}, and
\isotope[138]{Ba}(\isotope[18]{O}, 2n)\isotope[154]{Gd}
have been measured and compared with theoretical calculations with PACE4.
\end{abstract}


\end{frontmatter}

\section{Introduction}
\label{introduction}

\begin{figure}[ht]
    \centering
    \includegraphics[width=0.5\textwidth]{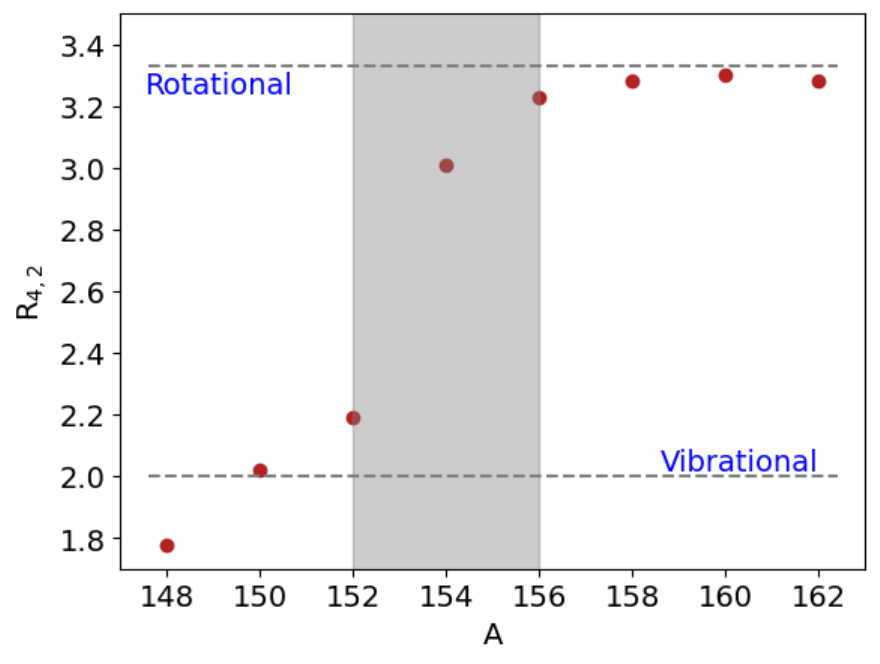}
    \caption{Degree of deformation as a function of mass number in Gd isotopes. $R_{4,2}$ is the ratio between the excitation energies of the first 4$^+$ and 2$^+$ states. The gray area indicates the transitional region in Gd, to which the three isotopes studied in the present work belong.}
    \label{fig:Gd_transition}
\end{figure}

The rare earth region of the nuclear chart has been a major focus of both experimental
and theoretical studies over the past years, mostly due to the onset of deformation and
the collective nature of its isotopes. In particular, a range of deformed nuclei can be
observed, ranging from the transitional Gd to the well-deformed Yb and Hf~\cite{zyriliou22,vasileiou23}, and the $\gamma$-soft
Pt isotopes. One special feature of even-even rare earths is the nature of the $K^{\pi} = 0^+$
states, which albeit been studied extensively in recent
years~\cite{0+_first,0+_second,0+_third,0+_forth,Gupta2015}, still remains a difficult
puzzle. Some of the most important open questions regarding these states are the existence
of many $K^{\pi}$ = 0$^+$ states below the pairing gap and the major variations in the
collectivity of excited levels, both in the same nucleus, as well as in close isotopic
chains. From this group of nuclei, the neutron-rich Gd isotopes have been one of the main
areas of research~\cite{PhysRevC.66.051305,PhysRevC.100.044324}, largely due to their
position between the $N\leqslant 88$ (vibrational) and $N\geqslant 90$ (rotational) nuclei
(Fig. \ref{fig:Gd_transition}).

Moreover, Gd isotopes have been the focus of many neutron-capture reactions, owing to Gd's
exceptionally high thermal neutron cross section among all stable elements~\cite{PhysRevC.88.034317,
PhysRevC.84.014306, PhysRevC.87.044609}. These reactions offer insights into the nuclear
level density and photon strength function. Despite the intense study of the nuclear
community, however, various spectroscopic pieces of information are still absent, including
several branching ratios in the deformed states and relative cross sections to the various
Gd channels populated in nuclear reactions.

A successful way to access the nuclei in the rare earth region, including Gd, is by using
fusion-evaporation reactions. The interest in such reactions became stronger in the 1960's
and 1970's, a period overlapping with the appearance of new accelerator facilities that
provided the necessary beams. An advantage of fusion-evaporation reactions lies in the significant
transfer of angular momentum to the various reaction channels, leading to the population
of many high-spin excited states. The study of these states as they decay to lower energies
provides information on the interplay between single-particle and collective degrees of freedom,
as well as, on band terminations~\cite{PhysRevC.75.064314,PhysRevC.72.024317}. Furthermore,
arranging these states in increasing spin order can reveal nuclear structure phenomena,
including shape evolution.

At lower energies, especially close to the Coulomb barrier, fusion reactions and nuclear
structure were shown to exhibit an interesting link (see e.g.~\cite{PhysRevC.75.064314,
PhysRevC.72.024317}). This fact is further emphasized in two review publications in which
the developments in heavy-nuclei fusion reactions are summarised~\cite{BACK2014}, while
some recent experimental results are highlighted~\cite{Montagnoli2017}, especially in the
region below and above the Coulomb barrier.

Fusion reactions have also been on the spotlight of numerous theoretical studies. A detailed
theoretical summary of quantum tunneling present at energies below and up to the Coulomb barrier
is given in~\cite{HAGINO2012}. The main point of this work, as well as of the previously mentioned
reviews, is that even though one-dimensional potential approaches (e.g.~\cite{PhysRevLett.39.265})
are sufficient to reproduce experimental data above the barrier, they severely underestimate the
cross section under it. There, coupling to low-energy collective modes enhances the fusion cross
section and coupled-cluster (CC) calculations are needed to reproduce the
data~\cite{BROGLIA198334,DASSO1983221,DASSO1983381,DASSO1985495}. Thus, experimental input of
excitation functions can not only provide additional benchmarking for the theory, but also allow
for the extraction of fusion barrier distributions. The latter are crucial for identifying the
nature of coupling responsible for sub-barrier fusion enhancement.

This work presents the first attempt to populate the \isotope[152,153,154]{Gd} isotopes by the fusion reaction
\isotope[18]{O}+\isotope[138]{Ba} and the subsequent evaporation to the various Gd channels
in the center-of-mass energy range of 58-64~MeV. The results of the analysis described in the
next sections include several new and updated branching ratios in the deformed bands of
\isotope[152,153]{Gd}, including the $K^{\pi}$ = 0$^+$ band and high-spin states. Additionally,
relative cross sections to the various Gd channels have been measured for the first time and
are compared with theoretical calculations.

\section{Experimental setup}
\label{Experimental setup}

The experiment was conducted at the 9~MV Tandem Accelerator of the Horia Hulubei National Institute
of Physics and Nuclear Engineering (IFIN-HH) in Magurele, Romania. \isotope[18]{O} beams at four
lab energies between 61 and 67 MeV have been used to impinge the \isotope[138]{Ba} target for a
total run time of approximately 63 hours.

\subsection{The target}

The manufacturing of the Ba target poses certain difficulties, since Barium is a metal that oxidizes
rapidly. To slow down the oxidation process, and protect the integrity
of the measurement, both sides of the target were covered with thin layers of Au. The whole
manufacturing process was performed at the Target Preparation Laboratory of IFIN-HH~\cite{refId0}
in argon atmosphere. The natural Ba target (see Table \ref{tab:Ba_composition} for the relative abundances) was created through the metallothermic reduction reaction of BaCO$_3$ with La metal powder. To create it,
a Au foil of 4.88~mg/cm$^2$ thickness was glued to the front of the target frame, 4~cm above the
Quorum Technologies E6700 bench evaporator tantalum boat. The boat was connected to the evaporator
via two high current cables.

\begin{table}[ht]
    \caption{Isotopic abundances of the metallic Ba target~\cite{Abundances}}
    \centering
    \begin{tabular}{c|c}
    \hline\hline
    Isotope & Abundance [$\%$] \\
    \hline
      \isotope[138]{Ba}   & 71.70(29) \\
       $^{137}$Ba  & 11.23(23) \\
       $^{136}$Ba & 7.85(24) \\
       $^{135}$Ba & 6.59(10) \\
       $^{134}$Ba & 2.42(15) \\
       Rest & 0.210(14) \\
       \hline\hline
    \end{tabular}
    \label{tab:Ba_composition}
\end{table}

At first, a low current was passed through the cables and it was slowly increased, so that the
desired temperature for the evaporation was reached. This process was continued, until the
2~mg/cm$^2$ thickness of the Ba layer was achieved. Finally, the other side of the target was
covered with a 0.5~mg/cm$^2$ Au foil. The latter was deposited with a Ta basket, placed 9~cm above
the substrate. The measurement of the 4.88~mg/cm$^2$ thickness of the Au layer was performed by
weighing on a precision scale, whereas for the other two foils, their values were determined by
calculating the thickness of the initially used materials.

\subsection{Detector array}

\begin{figure}[ht]
     \centering
     \begin{subfigure}[b]{0.5\textwidth}
         \centering
         \includegraphics[width=\textwidth]{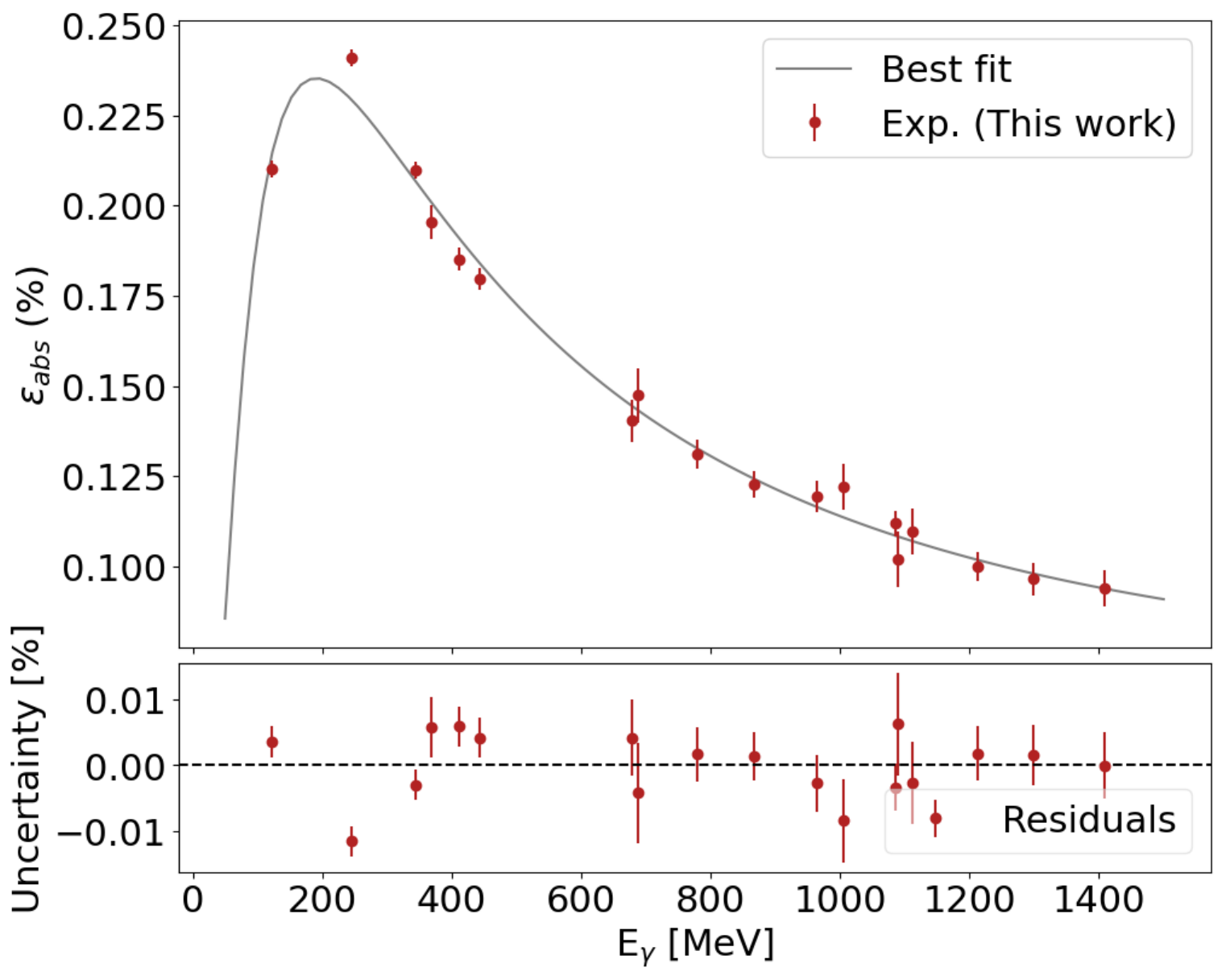}
         \caption{}
         \label{fig:det_3}
     \end{subfigure}
     \hfill
     \begin{subfigure}[b]{0.4\textwidth}
         \centering
         \includegraphics[width=\textwidth]{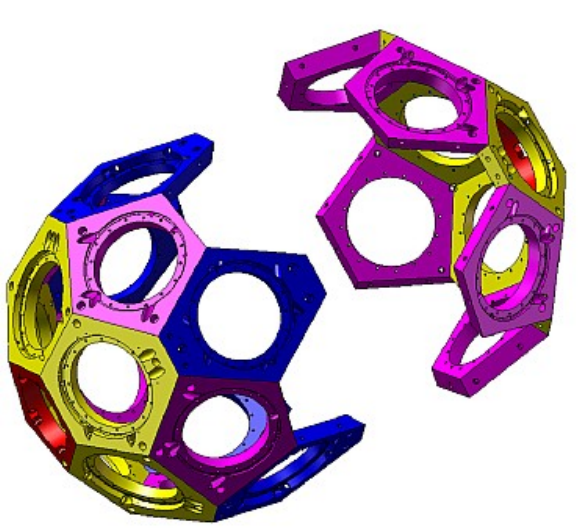}
         \caption{}
         \label{fig:rings}
     \end{subfigure}
          \caption{(a) Typical absolute efficiency curve for detector No. 3, positioned at the 37$^{\circ}$ ring. The subplot on the bottom indicates the uncertainty of the fit. (b) Illustration of the five rings of the ROSPHERE setup, placed at 37$^{\circ}$, 70$^{\circ}$, 90$^{\circ}$, 110$^{\circ}$ and 143$^{\circ}$. Figure adopted from \cite{BUCURESCU20161}.}
\end{figure}

For this experiment, the ROSPHERE array~\cite{BUCURESCU20161} was employed, in a configuration
featuring 15 HPGe detectors and 10 LaBr$_3$(Ce) scintillators. The events from the latter were not
used in the event trigger, to obtain sufficient statistics. The HPGe detectors were distributed
in sets of five across three rings at 37$^{\circ}$, 90$^{\circ}$ and 143$^{\circ}$, respectively. A typical efficiency curve for the Detector No. 3, along with an illustration
of the ROSPHERE setup, are presented in Figs. (\ref{fig:det_3}) and (\ref{fig:rings}). All detectors
exhibited a similar behavior. The efficiency measurements and the energy calibration were performed
with a \isotope[152]{Eu} point calibration source.

\section{Experimental results}
\label{Experimental results}

In the present section the results of the data analysis regarding the energy levels in
\isotope[152,153,154]{Gd}, which were populated by the heavy-ion fusion-evaporation reactions
\begin{equation*}
\begin{split}
\isotope[18]{O} + \isotope[138]{Ba} \rightarrow \Big[\isotope[156]{Gd}\Big]
\rightarrow \isotope[152]{Gd} + 4n \\
\rightarrow \isotope[153]{Gd} + 3n \\
\rightarrow \isotope[154]{Gd} + 2n
\end{split}
\end{equation*}
are presented in some detail, focusing on branching ratios and relative cross sections.

\subsection{Branching ratios}

If an energy level $E_{lvl}$ is de-populated by more than one $\gamma$-ray emission, then
the relative intensities of those de-populations have to be determined to provide information
on the transition matrix elements governing the particular decays. In the present work,
the branching ratios, expressed by the relation:
\begin{equation}
    b.r. (E_{lvl}) = \frac{Y_i}{\sum_j Y_j}
    \label{eq:BR}
\end{equation}
are deduced from the experimental data, using $\gamma$-$\gamma$ coincidence methods. In Eq. (\ref{eq:BR}), $Y_i$ is the yield of one of the
emitted $\gamma$-rays and $\sum_j Y_j$ is the sum of the yields for all the possible $\gamma$-ray
emissions from that level.

\begin{figure}[ht]
\centering
\begin{subfigure}[b]{0.49\textwidth}
    \centering
   \includegraphics[width=\textwidth]{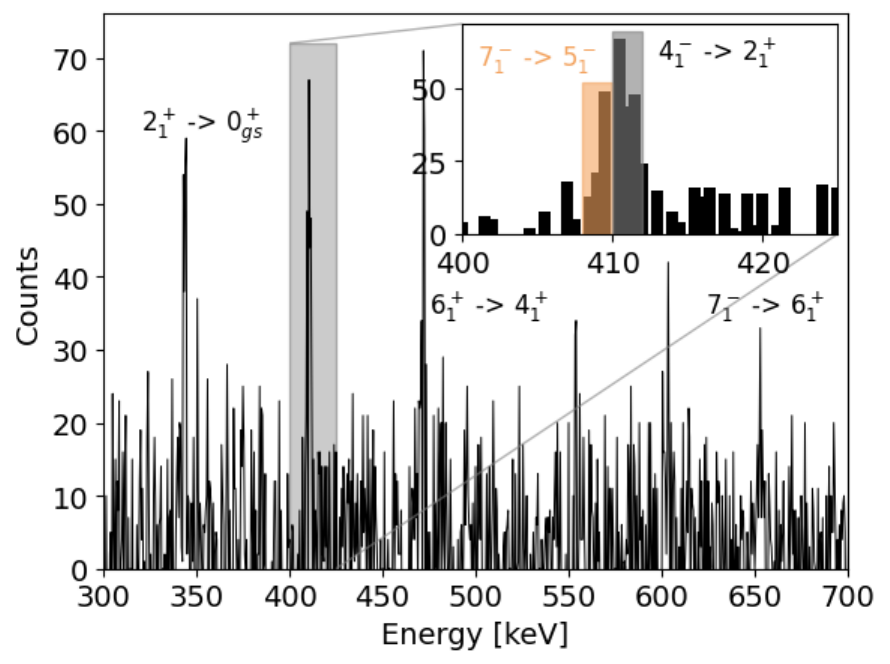}
   \caption{}
   \label{fig:Ng1}
\end{subfigure}
\begin{subfigure}[b]{0.49\textwidth}
    \centering
   \includegraphics[width=\textwidth]{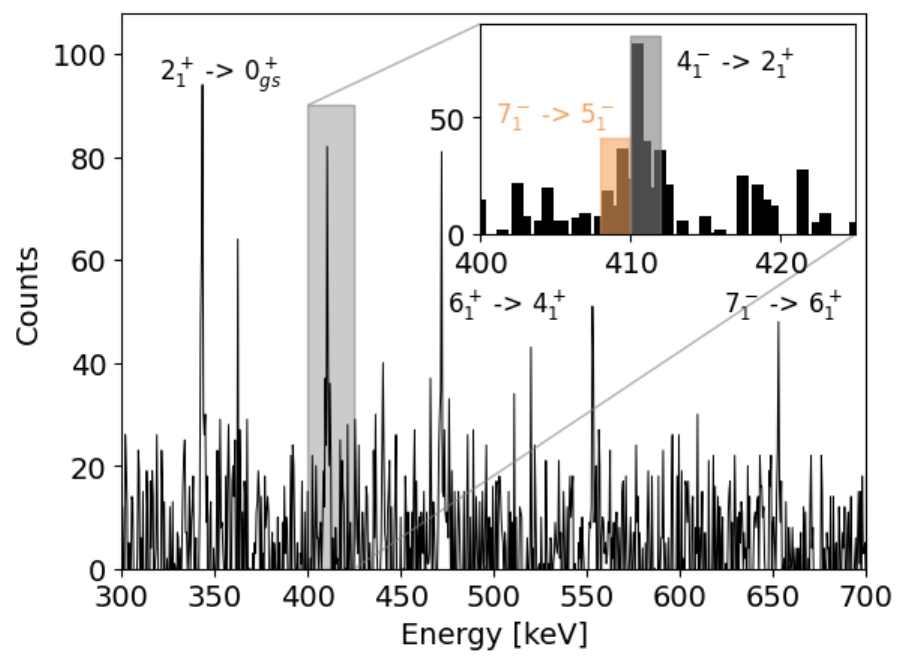}
   \caption{}
   \label{fig:Ng2}
\end{subfigure}
\caption{(a) Projection of the $\gamma$-$\gamma$ spectrum of $^{152}$Gd, gated on the 9$^-_1$ $\rightarrow$ 7$^-_1$ transition with $E_\gamma$ = 451.1 keV. (b) Projection of the $\gamma$-$\gamma$ spectrum of $^{152}$Gd, gated on the 8$^+_2$ $\rightarrow$ 7$^-_1$ with $E_\gamma$ = 258 keV. The two zoomed regions depict the partially overlapped peaks at $\approx$ 410 keV.}
\label{fig:spectra}
\end{figure}

\begin{figure}[p]
    \centering
    \includegraphics[width=0.7\linewidth, angle = 270]{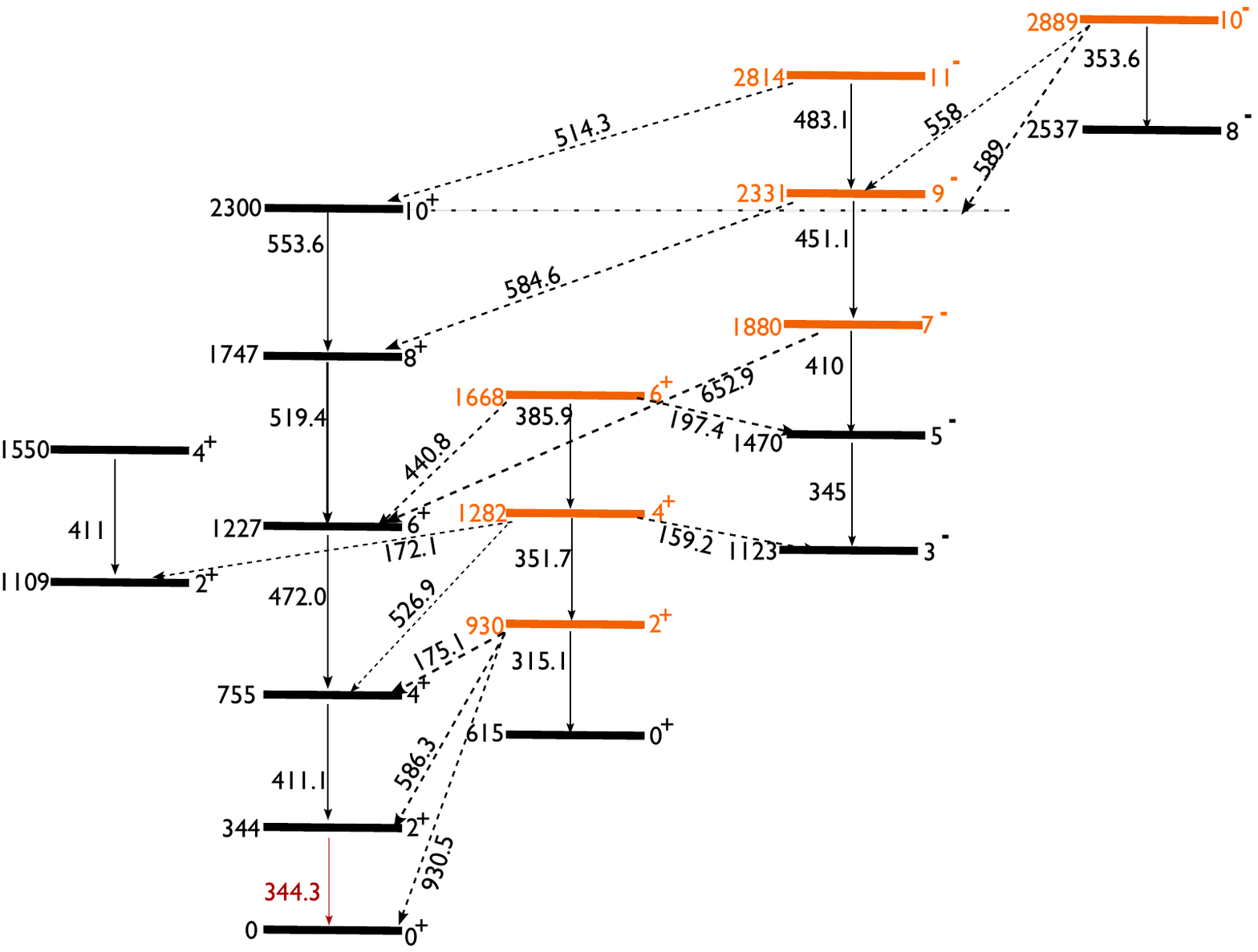}
    \caption{Partial level scheme of \isotope[152]{Gd}. The assignment of the bands from left to right is Band 10 ($K^\pi$ = 2$^+$), Band 1 ($K^\pi$ = 0$^+$ gs band), Band 2 ($K^\pi$ = 0$^+$ quasi $\beta$), Band 3 and Band 4. The red arrow ($2^+\rightarrow 0^+$; 344.3~keV) indicates the transition used for the extraction of the relative cross sections. Orange levels indicate the branching ratios measured in this work. All level and transition energies are given in keV.}
    \label{fig:152Gd}
\end{figure}

\begin{figure}[p]
    \centering
    \includegraphics[width=0.5\linewidth, angle = 270]{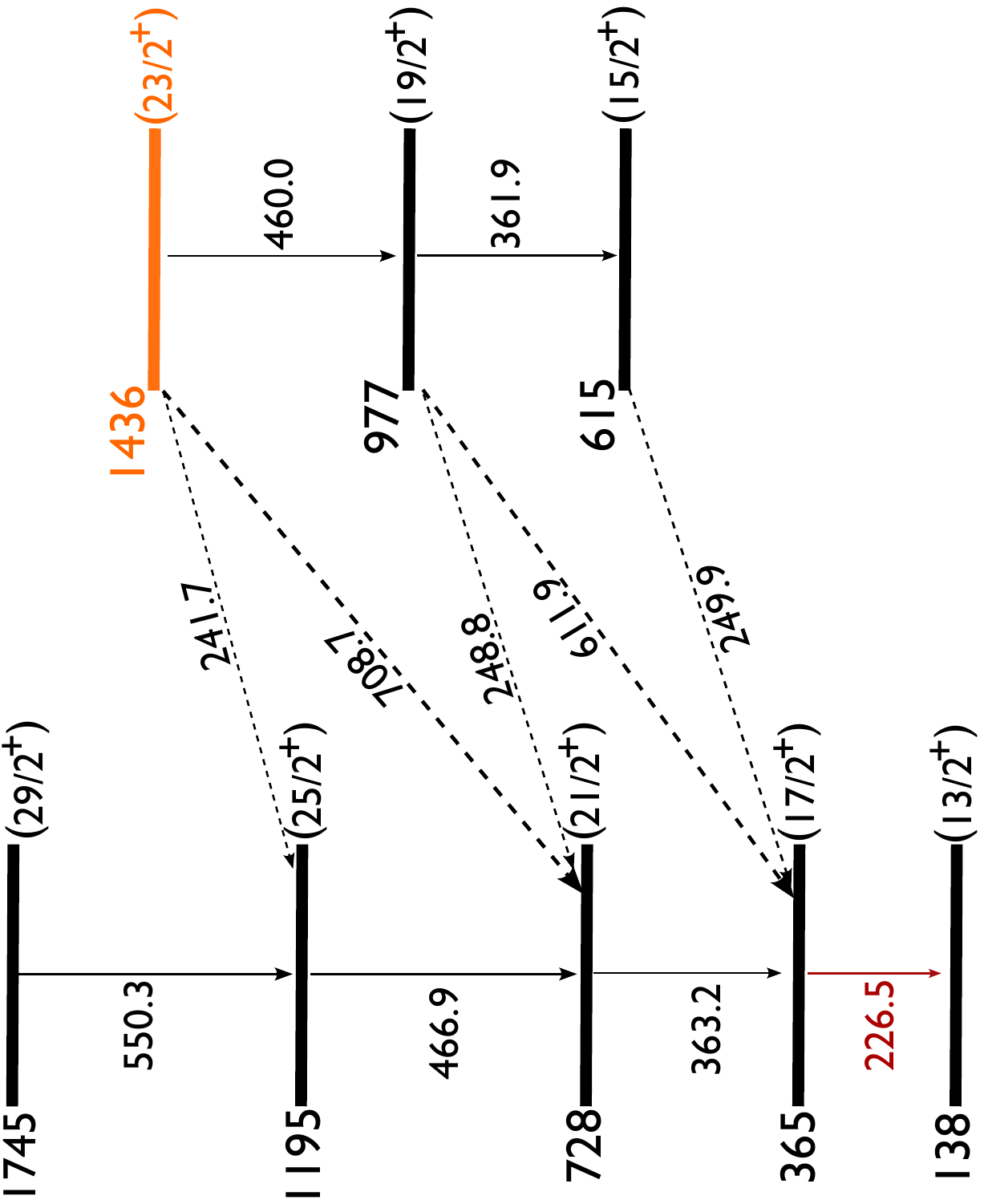}
    \caption{Partial level scheme of \isotope[153]{Gd}. The levels on the left are non-band levels, while the ones on the right belong to Band 2. The red arrow  ($17/2^+\rightarrow 13/2^+$; 226.5~keV) indicates the transition used for the extraction of the relative cross sections. Orange levels indicate the branching
    ratios measured in the present work. All level and transition energies are given in keV.}
    \label{fig:153Gd}
\end{figure}

\begin{figure}[p]
    \centering
    \includegraphics[width=0.5\linewidth, angle = 270]{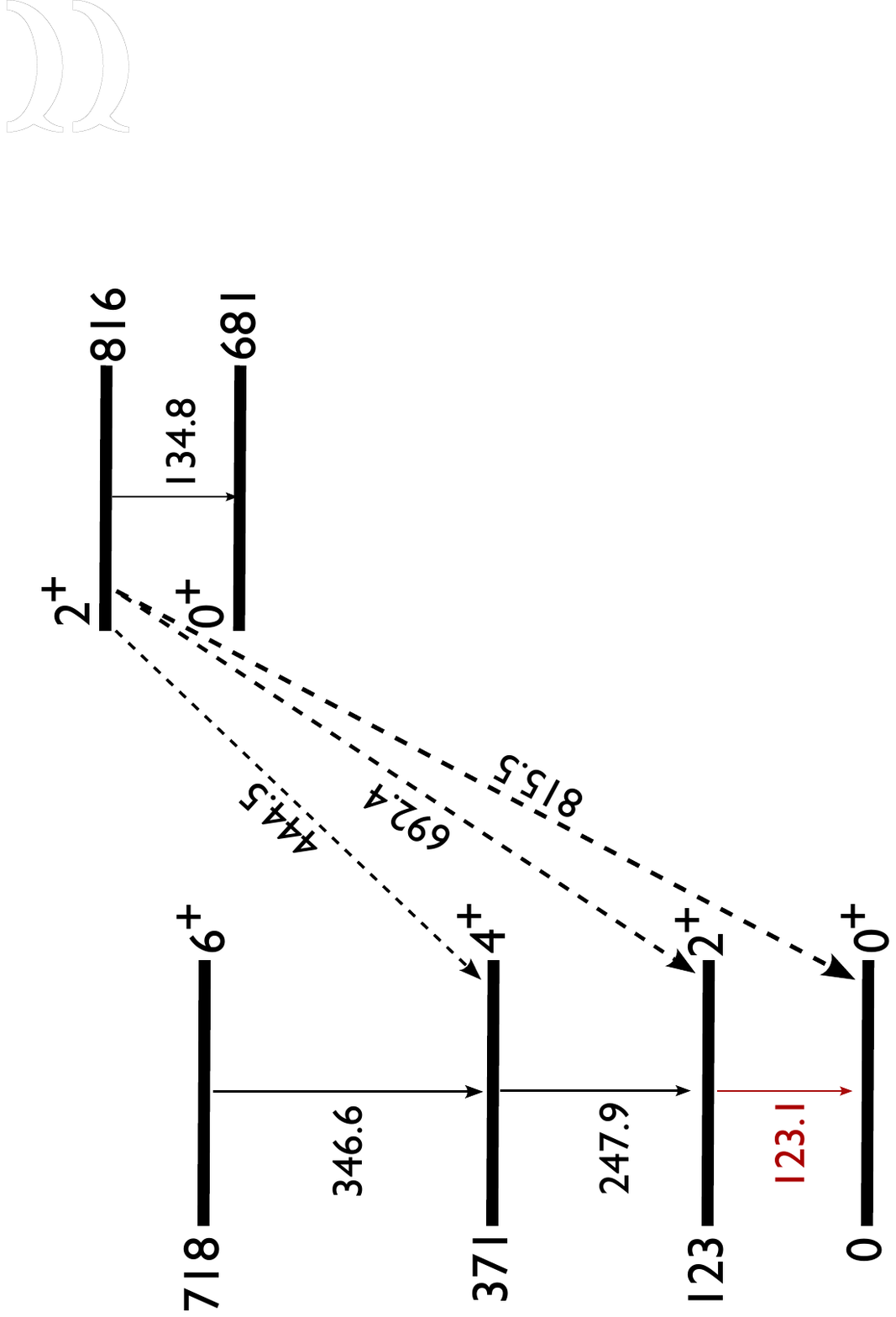}
    \caption{Partial level scheme of \isotope[154]{Gd}. The assignment of the Bands from left to right are Band 1 ($K^\pi$ = 0$^+$ gs) and Band 2. The red arrow  ($2^+\rightarrow 0^+$; 123.1~keV) indicates the transition used for the extraction of the relative cross sections. No branching ratios were measured for this isotope. All level and transition energies are given in keV.}
    \label{fig:154Gd}
\end{figure}

The experimentally determined values for the cases of \isotope[152,153]{Gd}, along with literature
values, are presented in Tables~\ref{tab:br152Gd} and \ref{tab:br153Gd}, respectively. The empty entries correspond to transitions not found in the spectra. A visualisation
of the relevant cascades is also given in Figs.~\ref{fig:152Gd}-\ref{fig:154Gd}.

To clear the spectra from the majority of the contaminant peaks, an energy gate was placed on the
$\gamma$-transition that feeds the de-exciting levels.
The branching of the 7$^-_1$ state in \isotope[152]{Gd} was studied with the application of two
different energy gates, one on the
$9^-_1 \rightarrow 7^-_1$ transition (with $E_\gamma = 451.1$~keV) and another on the
$8^+_2 \rightarrow 7^-_1$ transition (with $E_\gamma = 258$~keV). Graphical details can be seen in Fig.~\ref{fig:spectra}. The results for the first case produce a branching ratio of 0.42(16) for the $7^-_1 \rightarrow 5^-_1$
de-excitation and 0.58(19) for the transition $7^-_1 \rightarrow 6^+_1$, after taking into account the
relative efficiencies.  For the second energy gate, the final branching ratios were found equal to
0.46(20) for the transition $7^-_1 \rightarrow 5^-_1$, and 0.54(23) for the transition $7^-_1 \rightarrow 6^+_1$. These two different set of values have an excellent agreement with each other and have not been measured before in literature. The final values listed
in Table~\ref{tab:br152Gd} are the averaged results from both approaches.

The rest of the measurements of branching ratios in $^{152,153}$Gd from this work were reported previously. The two sets of data agree within 1 $\sigma$ with each other, with the exception being the ratios of the 2$^+_2$ and 6$^+_2$ levels in $^{152}$Gd. No comparison can be made for the 11$^-_1$, 10$^-_1$ and 9$^-_1$ levels in $^{152}$Gd, since the literature values are given without errors.

\begin{table}[ht]
    \begin{center}
    \caption{Experimentally deduced gamma energies and branching ratios for decays in \isotope[152]{Gd} compared to values quoted as found in literature. A multipolarity within square brackets indicates that the value has been deduced from the spin-parity combination of the initial and final state, but has not yet been experimentally determined. If a multipolarity is listed as D, it means that it is a dipole one, but the order has not been deduced. Similarly, Q refers to a quadrupole multipolarity.}
    \begin{tabular}{c|cc|cc|c}
    \hline\hline
    transition & \multicolumn{2}{c|}{E$_{\gamma}$ [keV]} & \multicolumn{2}{c|}{branching ratio} & multipolarity \\
    & this work & from ref.~\cite{MARTIN20131497} &  this work & from ref.~\cite{MARTIN20131497} & from ref.~\cite{MARTIN20131497} \\
    \hline
    $4^+_2$  $\rightarrow$ $4^+_1$  &527.00(5)&  526.88(5) & 0.47(5) & 0.512(15) & E0 + M1 + E2\\
    $4^+_2$  $\rightarrow$ $2^+_2$  &351.36(5)&  351.69(4) & 0.53(7) & 0.452(14) & E2\\
    $4^+_2$  $\rightarrow$ $2^+_3$  &--&   172.1(4) & -- & 0.017(8) & [E2]\\
    $4^+_2$  $\rightarrow$ $3^-_1$  &--& 159.16(16) & -- & 0.0184(21) & [E1]\\ \hline
    $2^+_2$  $\rightarrow$ $2^+_1$  &585.47(6)& 586.2648(26) & 0.70(9) & 0.800(9) & E0 + M1 + E2\\
    $2^+_2$  $\rightarrow$ $0^+_{gs}$ &930.02(25)& 930.50(4) & 0.15(4) & 0.128(3) & (E2) \\
    $2^+_2$  $\rightarrow$ $0^+_2$  &314.85(16)&  315.11(3) & 0.15(5) & 0.0703(15) & E2 \\
    $2^+_2$  $\rightarrow$ $4^+_1$  &--&  175.09(3) & -- & 0.0022(5) & [E2] \\ \hline
    $11^-_1$ $\rightarrow$ $10^+_1$ &514.24(3)&  514.3 & 0.65(10) & 0.65 & D \\
    $11^-_1$ $\rightarrow$ $9^-_1$  &483.20(19)&  483.1 & 0.35(7) & 0.35 & E2 \\ \hline
    $10^-_1$ $\rightarrow$ $10^+_1$ &590.03(18)&  589.9 & 0.65(19) & 0.62 & -- \\
    $10^-_1$ $\rightarrow$ $8^-_1$  &352.41(18)&  353.6 & 0.35(15) & 0.38 & -- \\
    $10^-_1$ $\rightarrow$ $9^-_1$  &--&  558.0 & -- & -- & -- \\ \hline
    $9^-_1$  $\rightarrow$ $7^-_1$  &451.80(56)&  451.1 & 0.86(14) & 0.78 & E2 \\
    $9^-_1$  $\rightarrow$ $8^+_1$  &584.73(6)&  584.6 & 0.14(4) & 0.22 & D \\
    \hline
    $6^+_2$  $\rightarrow$ $4^+_2$  &385.73(3)&  385.9(1) & 0.50(7) & 0.83(5) & E2 \\
    $6^+_2$  $\rightarrow$ $6^+_1$  &440.69(11)&  440.8(2) & 0.22(6) & 0.10(4) & (M1 + E2) \\
    $6^+_2$  $\rightarrow$ $5^-_1$  &197.14(7)&  197.4(3) & 0.28(8) & 0.07(1) & [E1] \\ \hline
    $7^-_1$  $\rightarrow$ $5^-_1$  &410.93(5)&  410.0    & 0.44(12) & -- & -- \\
    $7^-_1$  $\rightarrow$ $6^+_1$  &652.66(20)&  652.9(3) & 0.56(15) & -- & E1 \\
    \hline\hline
    \end{tabular}

    \raggedright
    \vspace{1em}
    \emph{}
    \label{tab:br152Gd}
    \end{center}
\end{table}

\begin{table}[ht]
    \begin{center}
    \caption{Experimentally deduced gamma energies and branching ratios for decays in \isotope[153]{Gd} compared to values found in literature. A spin-parity or multipolarity value in parenthesis means that there are experimental evidence to suggest it, but there is not enough certainty for the assignment.}
    \begin{tabular}{c|cc|cc|c}
    \hline\hline
    transition & \multicolumn{2}{c|}{E$_{\gamma}$ [keV]} & \multicolumn{2}{c|}{branching ratio} & multipolarity \\
    & this work & from ref.~\cite{NICA20201} &  this work & from ref.~\cite{NICA20201} & from ref.~\cite{NICA20201} \\
    \hline
    $\left(\frac{23}{2}^+\right)$ $\rightarrow$ $\left(\frac{19}{2}^+\right)$ &459.4(3)& 460.0(2) & 0.59(22)  & 0.53(4)  & E2 \\
    $\left(\frac{23}{2}^+\right)$ $\rightarrow$ $\left(\frac{21}{2}^+\right)$ &708.2(3)& 708.7(2) & 0.41(25)  & 0.36(3)  & D, Q \\
    $\left(\frac{23}{2}^+\right)$ $\rightarrow$ $\left(\frac{25}{2}^+\right)$ &--& 241.7(5) & -- & 0.111(8)  & (M1 + E2) \\
    \hline
    \hline
    \end{tabular}

    \raggedright
    \vspace{1em}
    \emph{}

    \label{tab:br153Gd}
    \end{center}
\end{table}

\subsection{Relative cross sections}

The total cross section for a reaction that directly populates the ground state of an isotope
can be expressed by the relation~\cite{psaltis19,vasileiou21}:
\begin{equation}
    \sigma = \frac{A}{N_A} \frac{Y}{\delta}
\end{equation}
where $A$ is the atomic mass of the target, $N_A$ is the Avogadro constant, $\delta$ is the target
thickness and $Y$ is the total yield of the reaction. If there are more than one $\gamma$ transitions
that populate the ground state, then the total Yield is equal to:
\begin{equation}\label{eq:yield}
    Y = \sum_i Y_i = \sum_i \frac{N_i}{N_f \epsilon_{abs}}
\end{equation}
where $N_i$ is the number of measured $\gamma$-rays, $N_f$ is the number of incident particles
on the target and $\epsilon_{abs}$ is the total efficiency of the setup, for the corresponding
energy of the $\gamma$-rays.

Considering the above, the relative cross section for two different reaction channels $\alpha$
and $\beta$ can be deduced via:
\begin{equation}
    \sigma_{rel} = \frac{N_i(\alpha) \epsilon_{abs}(\beta)}{N_i(\beta) \epsilon_{abs}(\alpha)}
\end{equation}

\begin{table}[tb]
    \caption{List of transitions used for the extraction of the relative cross sections.
    Information on the spins and parities of the transitions, as well as the gamma-ray energies are quoted as found in literature.}
    \centering
    \begin{tabular}{cccc} \hline
    \hline
    Isotope & Transition & E$_{\gamma}$ [keV] & Reference \\
    \hline
    \isotope[152]{Gd} & 2$^+_{1}$ $\rightarrow$ 0$^+_{gs}$ & 344.3 & \cite{MARTIN20131497} \\
    \isotope[153]{Gd} & $\left( \frac{17^+}{2} \right)$ $\rightarrow$ $\left( \frac{13^+}{2} \right)$ & 226.5(10) & \cite{NICA20201} \\
    \isotope[154]{Gd} & 2$^+_{1}$ $\rightarrow$ 0$^+_{gs}$ & 123.1(1) & \cite{REICH20092257} \\
    \hline
    \hline
    \end{tabular}
    \label{tab:Energies}
\end{table}

\begin{figure}[ht]
\centerline{\includegraphics[width=0.85\textwidth]{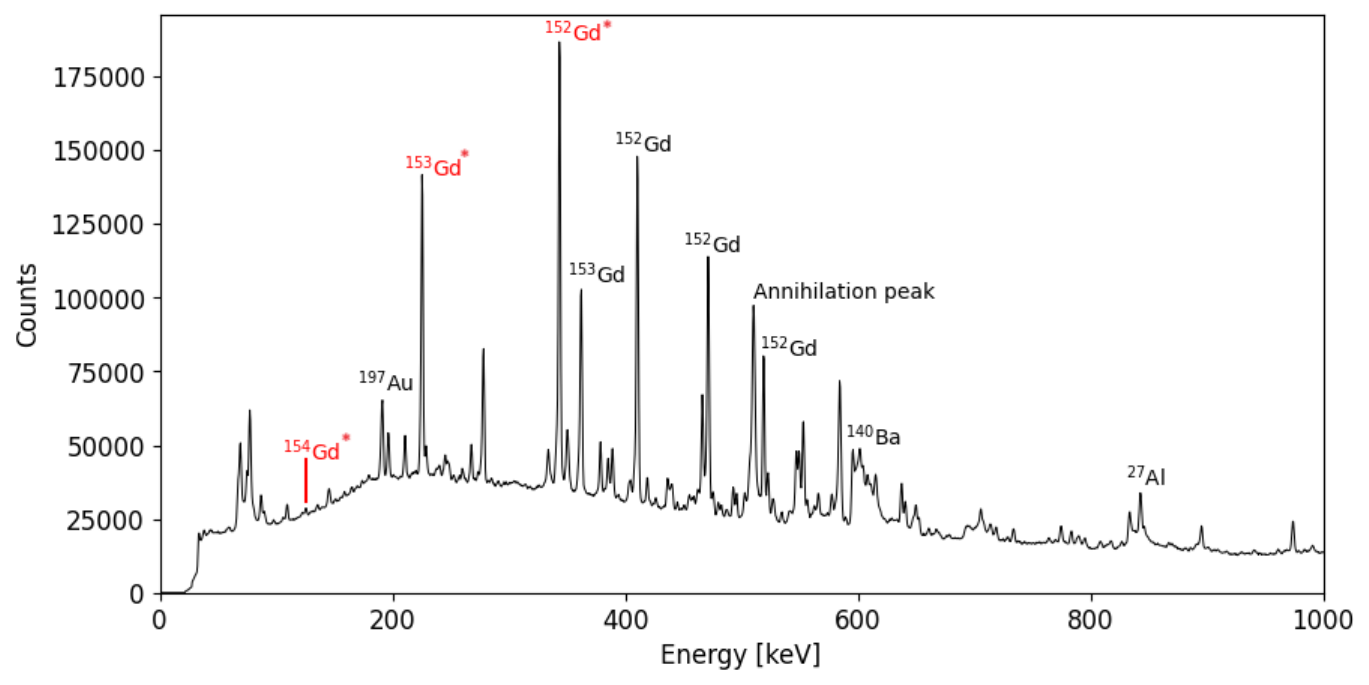}}
\caption{Total projection spectrum of the $\gamma$-$\gamma$ matrix for E$_{lab}$ = 65 MeV. The main transitions for \isotope[152-154]{Gd}, along with some contaminant peaks, are depicted. The values with an asterisk are used to indicate the peaks for the extraction of the relative cross sections.}
\label{fig:65MeV_spectrum}
\end{figure}

For the cases of \isotope[152]{Gd} and \isotope[154]{Gd}, the transitions of interest are the
pure (E2) $2^+_1 \rightarrow 0^+_{gs}$ with $E_{\gamma} = 344.3$~keV and 123.1~keV, respectively.
The situation is different for \isotope[153]{Gd}, since the lowest measured state is the
$J^{\pi} = \left(\frac{13}{2}^+\right)$, positioned at 138.4~keV. No $\gamma$-rays appear to be
emitted from this level, and according to the latest evaluation in Nuclear Data Sheets for this isotope~\cite{NICA20201}, it is considered as an isomer. However, no other information, such as its
lifetime, is known. For this reason, the (E2) transition
$\left(\frac{17}{2}^+\right)$ $\rightarrow$ $\left(\frac{13}{2}^+\right)$
with $E_{\gamma} = 226.5$~keV
is treated as the one for the deduction of the relevant, to \isotope[153]{Gd}, cross section.
These peaks of interest, along with a few others and some important contaminants, are depicted
in Fig.~\ref{fig:65MeV_spectrum}. Table~\ref{tab:Energies} contains a list of the transitions
used for the cross section measurements.

\begin{figure}[ht]
\centering
\begin{subfigure}[b]{0.49\textwidth}
    \centering
   \includegraphics[width=\textwidth]{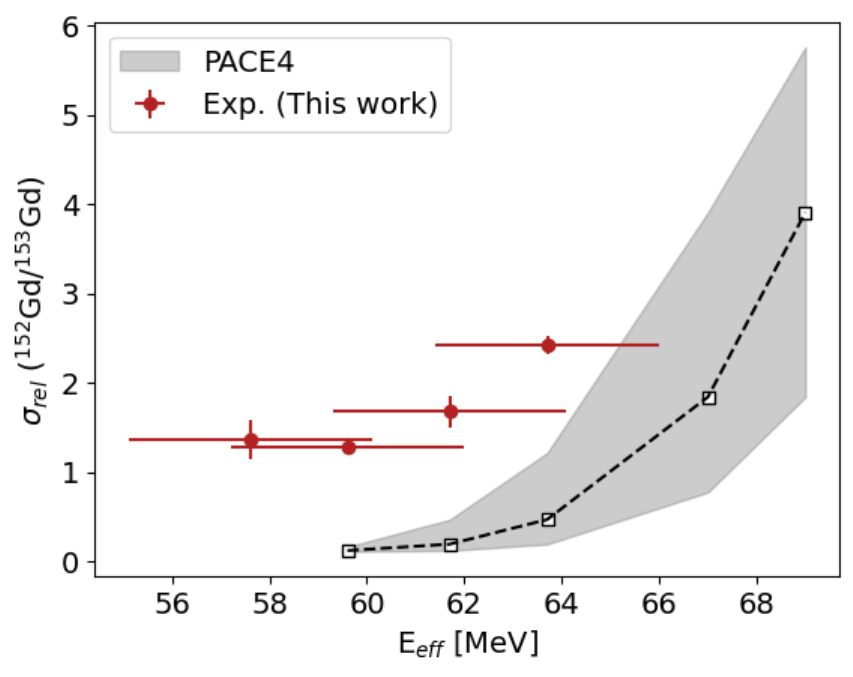}
   \caption{}
   \label{fig:RCS_152_153}
\end{subfigure}
\begin{subfigure}[b]{0.49\textwidth}
    \centering
   \includegraphics[width=\textwidth]{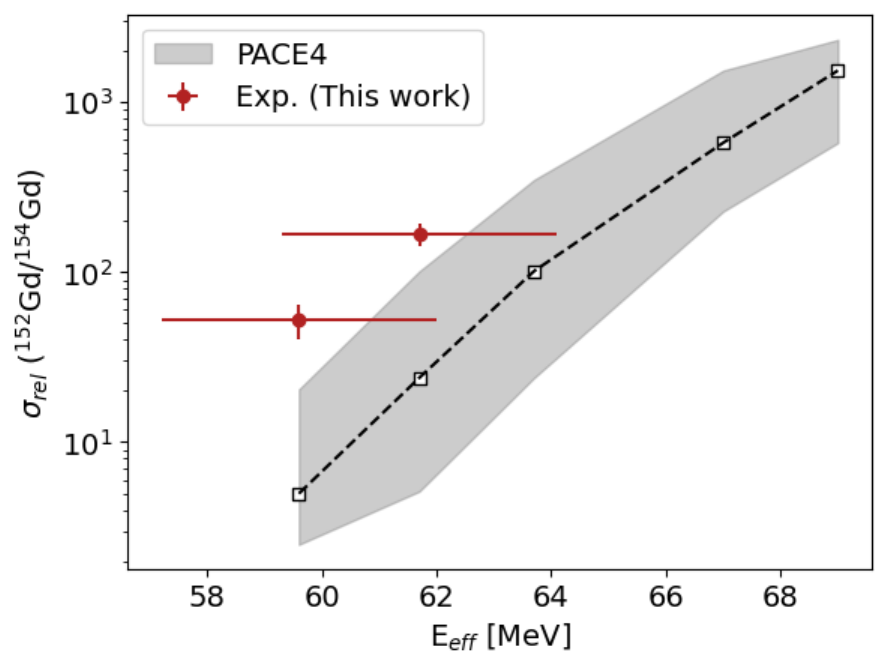}
   \caption{}
   \label{fig:RCS_152_154}
\end{subfigure}
\caption{(a) Relative cross sections of \isotope[152]{Gd} with respect to \isotope[153]{Gd}. (b): Relative cross sections of \isotope[152]{Gd} with respect to \isotope[154]{Gd}. The grey areas indicate the model predictions around a 2 MeV range.}
\label{fig:rcs}
\end{figure}

The results for the relative cross sections
(\isotope[152]{Gd}/\isotope[153]{Gd}) and
(\isotope[152]{Gd}/\isotope[154]{Gd})
are shown graphically in Fig.~\ref{fig:rcs}, along with theoretical calculations from the PACE4 code. The latter is a satellite code incorporated into the LISE++ network \cite{TARASOV2016185}.
Due to the presence of multiple isotopes in the metallic Ba target, the cross sections of the Gd
isotopes include contributions from reaction channels, other than the \isotope[18]{O} + \isotope[138]{Ba},
namely the channels following \isotope[18]{O} + \isotope[135-137]{Ba}. Such contributions are accounted for
in the calculated results, using the relative abundances (Table \ref{tab:rcs}). Note that the code was not able to give results for the lab energy of 61 MeV. To better take into
account the predictions of the model, results were calculated for a range of 2~MeV around the corresponding
energy at the center of the Ba foil $E_{eff}$. The latter was deduced from the relation:
\begin{equation}
    E_{eff} = E_{lab} - \frac{\Delta E}{2}
\end{equation}
where $\Delta E$ is the energy loss of beam inside the Ba foil, calculated with the Physical Calculator
tool of the LISE++ program~\cite{TARASOV2016185}.

\begin{table}[ht]
    \caption{Relative cross sections ($\sigma_{rel}$) of \isotope[152]{Gd} with respect to $^{153,154}$Gd for the lab energy range of 61-67 MeV. The two extra values of E$_{eff}$ were added to demonstrate the behavior of the PACE4 algorithm close to the Coulomb barrier. The uncertainties limits in the calculated cross sections are obtained from the range of predictions of 2 MeV around the E$_{eff}$ of each case. E$_{lab}$ is the energy of the beam in the laboratory frame and E$_{eff}$ the energy in the center of the Ba layer, approximately where the reaction takes place.}
    \centering
    \begin{tabular}{cccccc} \hline
    \hline
    \multicolumn{1}{c}{$E_{lab}$} & \multicolumn{1}{c}{$E_{eff}$} & \multirow{2}{*}{$\sigma^{^{152}Gd/^{153}Gd}_{rel,exp}$} & \multirow{2}{*}{$\sigma^{^{152}Gd/^{154}Gd}_{rel,exp}$} & \multirow{2}{*}{$\sigma^{^{152}Gd/^{153}Gd}_{rel,PACE}$} & \multirow{2}{*}{$\sigma^{^{152}Gd/^{154}Gd}_{rel,PACE}$}  \\
    \multicolumn{1}{c}{$[$MeV$]$} & \multicolumn{1}{c}{$[$MeV$]$} & & & &  \\ \hline
    -- & 69 & -- & -- & 3.91$_{-2.07}^{+1.85}$  & 1535.19$_{-958.21}^{+793.28}$  \\
    -- & 67 & -- & -- & 1.84$_{-1.07}^{+2.07}$ & 576.98$_{-349.04}^{+958.21}$  \\
    67 & 63.7(23) & 2.42(10) & -- & 0.47$_{-0.28}^{+0.75}$ & 101.33$_{-77.45}^{+247.97}$ \\
    65 & 61.7(24) & 1.68(18) & 168.2(256) & 0.19$_{-0.07}^{+0.28}$ & 23.88$_{-18.74}^{+77.45}$  \\
    63 & 59.6(24) & 1.28(9) & 52.0(119) & 0.12$_{-0.01}^{+0.04}$ & 4.97$_{-2.48}^{+15.46}$  \\
    61 & 57.6(25) & 1.37(22) & -- & -- & --  \\
    \hline
    \hline
    \end{tabular}
    \label{tab:rcs}
\end{table}

As it can be seen from Fig.~\ref{fig:rcs}, PACE4 data have a similar trend, but they significantly
differ from the experimental values. This is somewhat expected however: PACE4 is a modified version
of the JULIAN code, which is based on the Bass model~\cite{PhysRevLett.39.265} for the determination
of the nuclear potential, during the fusion of two nuclei. However, as it is noted by Bass, and later
by Reisdorf~\cite{Reisdorf_1994}, the results from the Bass model are in excellent agreement with
experimental data only for energies above the Coulomb barrier and as long as the fission channel is
small ($< 5\%$). Meanwhile, various experimental attempts have shown that this model tends to
underestimate the cross section values below the barrier~\cite{Reisdorf_1994}. For the fusion reaction
$\isotope[18]{O} + \isotope[138]{Ba} \rightarrow \Big[\isotope[156]{Gd}\Big]$, the Coulomb barrier
is approximately placed at 66.2~MeV \cite{Catkin}, which is notably higher than all of the experimental $E_{eff}$ values. However,
as it was expected and later observed, the deviation between the experimental and the computed data
becomes smaller, as the difference between the barrier and $E_{eff}$ decreases.

\section{Discussion and conclusions}
\label{Discussion}

The relative cross sections to the three most dominant fusion channels leading to
\isotope[152-154]{Gd} were experimentally determined by measuring the transitions
directly feeding the ground state, with the exception of \isotope[153]{Gd}. The decay
scheme in the latter stops at the $\frac{13}{2}^+$ level positioned at 138.4(17) keV (Fig.~\ref{fig:153Gd}), thus limiting the relevant analysis. Out of the main fusion channels, the one to \isotope[152]{Gd}
was found to be the dominant at all four beam energies. The experimentally deduced
relative cross section results were compared with theoretical calculations from PACE4,
showing a strong deviation. This is a rather expected result, as PACE4 is known to be
fine-tuned for much higher energies than the Coulomb barrier. The present experimental
study seems to be in line with that.

From the nuclear structure point of view, several new and re-measured branching ratios in the deformed bands of \isotope[152,153]{Gd} were determined, using $\gamma$-$\gamma$ coincidence
techniques and compared with literature values from~\cite{MARTIN20131497,NICA20201}.
The results for the energy levels $J^{\pi}$ = $\left(\frac{23}{2}^+\right)$ in
\isotope[153]{Gd} and $J^{\pi}$ = 4$^+_2$, 2$^+_2$, 11$^-_1$, 10$^-_1$ and 9$^-_1$
in \isotope[152]{Gd} were re-measured and found to be in a very good agreement with existing values. It is worth mentioning that the last three of the above levels are not associated
with any uncertainty in the values included in latest compilation of evaluated data~\cite{MARTIN20131497}, to which the present work provides experimental results
for the first time. In addition, a branching ratio value for the $J^{\pi}$ = 7$^-_1$
state in \isotope[152]{Gd} has been deduced experimentally and is reported for the first time.

As the present work relies on having these Gd isotopes populated with a reaction used
for the first time in literature for that purpose, the experimental results complement
the global data sets rather nicely. More specifically, on the cross section part of
the work reported, relative cross section measurements have been produced for the first
time with the particular reaction mechanism. Although there is a justified disagreement
with the PACE4 calculations, the same trend can be seen in a number of previous publications,
with several being reported in \cite{Reisdorf_1994}. The sub-barrier experimental
dataset obtained from this work align with existing data, while serving as a future
starting point for new CC models. The latter can help confirm whether such effects
are based on the virtual or real excitation of low-lying excited states. Moreover,
these models can further be used to normalise the present relative cross sections.
The deduced excitation functions would then help to extract fusion barrier distributions,
shining more light into the mechanisms of sub-barrier fusion enhancement.

Additionally, the deduced branching ratios reflect the nuclear structure of the Gd
isotopes, which are conveniently placed at the transitional region between rotational
and vibrational nuclei. The re-measured values in this work are found to be in good
agreement with the literature data, while being measured with a different reaction.
This is particularly important, since several of these levels belong to $K^{\pi}$ = 0$^+$
bands. Finally, some new values are provided, which fill the nuclear puzzle in this
region, while also providing input for future theoretical modelling of angular
distributions of $\gamma$-rays following level depopulations (see e.g. \cite{Kulp}).

\section*{Acknowledgements}

The authors would like to thank M.~Efstathiou for assistance with the level scheme figures.
AC acknowledges support by the Hellenic Foundation for Research and Innovation (HFRI) and
the General Secretariat for Research and Technology (GSRT) under the HFRI PhD Fellowship Grant
(GA. No. 74117/2017). This research work was supported by the Hellenic Foundation for Research
and Innovation (HFRI) under the HFRI PhD Fellowship grant (GA No: 101742/2019) for AZ.
Partial support from ENSAR2 (EU/H2020 Research Infrastructure doi: 10.3030/654002) is kindly acknowledged.

\bibliographystyle{elsarticle-num}
\bibliography{bibliography}

\end{document}